\documentclass[reprint, amsmath,amssymb, aps]{revtex4-2}
\usepackage{graphicx}
\usepackage{dcolumn}%
\usepackage{bm}
\usepackage{relsize}
\usepackage{mathrsfs}
\usepackage{multirow}
\usepackage{textcomp}
\usepackage{textcase}
\usepackage{siunitx}
\usepackage{amsmath}
\usepackage{amssymb}
\usepackage{footnote}
\usepackage[dvipsnames]{xcolor}
\usepackage[utf8]{inputenc}

\DeclareUnicodeCharacter{2212}{-}

\newcommand{\old}[1]{}

\def\be{\begin{equation}}
	\def\ee{\end{equation}}
\def\bea{\begin{eqnarray}}
	\def\eea{\end{eqnarray}}

\begin{document}

\preprint{APS/123-QED}

\title{Relativistic equilibrium fluid configurations around rotating deformed compact objects}

\author{Shokoufe Faraji}
 \email{shohoufe.faraji@zarm.uni-bremen.de}
 \affiliation{%
 University of Bremen, Center of Applied Space Technology and Microgravity (ZARM), 28359 Germany
}%

\author{Audrey Trova}
 \email{audrey.trova@zarm.uni-bremen.de}
 \affiliation{%
 University of Bremen, Center of Applied Space Technology and Microgravity (ZARM), 28359 Germany
}%
\author{Hernando Quevedo}%
 \email{quevedo@nucleares.unam.mx}
\affiliation{Instituto de Ciencias Nucleares, Universidad Nacional Aut\'onoma de
M\'exico,
 AP 70543, Ciudad de M\'exico 04510, Mexico}
\affiliation{Dipartimento di Fisica and ICRANet, Universit\`a di Roma ``La
Sapienza",  I-00185 Roma, Italy}

\begin{abstract}
We investigate the physical properties of equilibrium sequences of non-self-gravitating surfaces that characterize thick disks  around a rotating deformed compact object described by a stationary generalization of the static q-metric. The spacetime corresponds to an exact solution of Einstein's field equations so that we can perform the analysis for arbitrary values of the quadrupole moment and rotation parameter. To study the properties of this disk model, we analyze  bounded trajectories in this spacetime. Further, we find that depending on the values of the parameters, we can have various disc structures that can  easily be distinguished from the static case and also from the Schwarzschild background.
We argue that this study may used to evaluate the rotation and quadrupole parameters of the central compact object.

\end{abstract}

\maketitle

\section{Introduction}

The gravitational field of astrophysical compact objects can be characterized by means of their multipole moments.  In Newtonian gravity, only the mass is a source of gravity and, therefore, all the multipole moments are determined by the mass distribution only. In the case of relativistic objects, there are two different sets of multipoles, namely, mass and angular momentum multipoles \cite{geroch1,geroch2,hansen}.  

From the point of view of their multipole structure, the simplest relativistic compact objects are black holes because they can be completely characterized by the lowest possible moments, i.e., by the mass monopole and the angular momentum dipole, which determine uniquely the corresponding Schwarzschild and Kerr spacetimes, respectively \cite{kerr}.
This is one of  the essential aspects of the black hole uniqueness theorems \cite{heusler}.

In the case of a non-rotating mass distribution, the next non-trivial multipole is the quadrupole, which describes the deviation of the mass distribution from spherical symmetry, but preserving the axial symmetry. In this case, no uniqueness theorems exist and so there are several possibilities to describe a spacetime of a mass with quadrupole. In \cite{aqs18}, it was established that there are  six different known solutions of Einstein equations that could be used to  describe the gravitational field of a mass with quadrupole. From all of them, we highlight  the quadrupolar metric ($\rm q$-metric) 
as the simplest generalization of the Schwarzschild spacetime, including a quadrupole \cite{quev11}. The q-metric can be obtained by applying a Zipoy-Voorhees transformation \cite{zipoy,voorhees} to the Schwarzschild metric and in spherical coordinates reads
\bea
ds^2 = && -\left(1-\frac{2m}{r}\right)^{1+q} dt^2 \nonumber \\
+ &&\left(1-\frac{2m}{r}\right)^{-1-q}\left(\frac{r^2-2mr+m^2\sin^2\theta }{r^2-2mr}\right)^{-q(2+q)} dr^2 \nonumber\\
+ &&\left(1-\frac{2m}{r}\right)^{-q}r^2(d\theta^2+\sin^2\theta d\phi^2)\ ,
\label{qmetric}
\eea
where $m$ and $\rm q$ are the mass and quadrupole parameters, respectively.

The $\rm q-$metric has been used to study the motion of test particles, accretion disks, black hole mimickers, shadows, interior and exterior counterparts, quasi periodic oscillations among others \cite{testmotion,qmetric4,abishev2016accretion,boshkayev2021luminosity,arrieta2020shadows,abishev2021approximate,2021A&A...654A.100F,2022Univ....8..195F,2021Univ....7..447F,2022MNRAS.tmp..870F}.
From these studies, it follows that the $\rm q-$metric satisfies all the physical conditions to describe the field of a deformed mass distribution with quadrupole. 

The next interesting physical aspect of compact objects is their rotation. Therefore, in the present work, as the next step of this work \cite{2021A&A...654A.100F}, we will consider a stationary generalization of the $\rm q$-metric that contains an additional parameter, corresponding to the dipole of the angular momentum
\cite{toktarbay2014stationary}.
We explore the applicability of the stationary $\rm q-$metric in  astrophysical situations, we study an axially symmetric thick accretion disk model, which is situated on the background of a rotating, deformed mass distribution. This hydrostatic equilibrium are strongly believed to form around  X-ray binaries, active galactic nuclei, and also in the central engine of gamma-ray bursts. The analytical thick disk model for an accretion disk is initially assumed to consist of an unmagnetised perfect fluid with constant angular momentum  \cite{1974AcA....24...45A,1978A&A....63..209K,1980AcA....30....1J,1980A&A....88...23P,1980ApJ...242..772A,1982MitAG..57...27P,1982ApJ...253..897P}. In this work we consider this analytical model and explain it briefly in Section \ref{sec:tori}.


The spacetime contains three independent multipoles, namely, mass monopole and quadrupole, and angular momentum dipole. We will show that the structure of accretion disks around the stationary $\rm q-$metric  depends explicitly on the values of all the independent parameters of the metric. In particular, we will see that the behavior of the accretion disks agrees with our physical expectations.

This work is organized as follows. In Section \ref{sec:sta}, we present the stationary $\rm q-$metric and comment on their main physical properties. In Section \ref{sec:tori}, we discuss the main theoretical aspects of the thick disk model and present the results in the spacetime described by the stationary $\rm q-$metric. In Section \ref{sec:con}, we discuss our results. In addition, the signature of the metric is set to be $(-,+,+,+)$ and we use geometrical units with $c=G=1$.


\section{The stationary \lowercase{q}-metric}
\label{sec:sta}

A stationary generalization of the static Zipoy-Vorhees spacetime is contained as a particular solution of the rotating Erez-Rosen solution and was first presented in \cite{quev90}. However, the physical meaning of the Zipoy-Vorhees parameter $\delta$ as a quadrupole parameter was first established and investigated only later on in \cite{quev11}. It was named $\rm q-$metric to emphasize the role of the $\rm q$ parameter as a quadrupole and as a source of naked singularities. The Ernst potential of the corresponding stationary generalization was presented in \cite{toktarbay2014stationary} and the explicit form of the metric was calculated in \cite{frutos2018relativistic}. In prolate spheroidal coordinates $(t,x,y,\phi)$, the corresponding line element can be written as
\begin{align}
   ds^2&= -f(dt-\omega d\phi)^2 \nonumber\\
   & + \frac{\sigma^2}{f}\left[e^{2\gamma}(x^2-y^2)\left(\frac{dx^2}{x^2-1}
   +\frac{dy^2}{1-y^2}\right)\right. \nonumber\\
   &\left.+(x^2-1)(1-y^2)d\phi^2\right],\
\end{align}
where $\sigma$ is a constant with the dimension of length and the metric functions  depend only on $x$ and $y$,
\begin{align}
    f&=\frac{A}{B},\nonumber\\
    \omega&=-2(a+\sigma \frac{C}{A}),\nonumber\\
    e^{2\gamma}&= \frac{1}{4}\left(1+\frac{m}{\sigma}\right)^2\frac{A}{(x^2-1)^{1+q}}\left[\frac{x^2-1}{x^2-y^2}\right]^{(1+q)^2}.\
\end{align} 
Here $\rm q$ represents the quadrupole parameter and $a$ is related to the angular momentum.
Besides
\begin{align}
    A=&a_{+} a_{-} + b_{+} b_{-},\\
    B=& a_{+}^2 + b_{+}^2,\nonumber\\
    C=&(x+1)^q\left[x(1-y^2)(\lambda + \eta)a_{+}+ y(x^2-1)(1-\lambda\eta)b_{+}\right],\nonumber\
\end{align} 
and
\begin{align}
    a_{\pm}=&(x\pm1)^q\left[x(1-\lambda \eta)\pm (1+\lambda \eta)\right],\nonumber\\
    b_{\pm}=&(x\pm1)^q\left[y(\lambda + \eta)\mp (\lambda - \eta)\right],\nonumber\\
    \lambda=& \alpha (x^2-1)^{-q}(x+y)^{2q},\nonumber\\
    \eta =& \alpha (x^2-1)^{-q}(x-y)^{2q}, \nonumber\\
    \alpha =& \frac{1}{a} (\sigma-m) \nonumber.\
\end{align}
The functions $f$ and $\omega$ are related to the twist scalar $\Omega$ through
\begin{align}
f^2\nabla{\omega}=\rho \phi \times \nabla{\Omega}.
\end{align}
The transformation between these prolate coordinates and cylindrical coordinates is given by
\begin{align}
\rho=\sigma \sqrt{(x^2-1)(1-y^2)}, \quad z=\sigma xy,
\end{align}
whereas the transformation to 
Schwarzschild-like coordinates reads 
\begin{align}\label{transf1}
 x =\frac{1}{\sigma}(r-m) \,, \quad  y= \cos\theta\,,
\end{align}
with
\begin{align}
\sigma&:= \frac{m(1-\alpha^2)}{(1+\alpha^2)},\\
a&=\frac{-2\sigma\alpha}{(1-\alpha^2)},\
\end{align}
so that 
\begin{equation}
    \sigma^2 = m^2 - a^2\ .
\end{equation}
Consequently, the condition $m \geq |a|$ must be satisfied. However, since the limiting case $m=|a|$ implies that $\mathcal{X}= \sigma =0$ and the constant $\sigma$ appears explicitly in the line element and metric functions, this limiting value could lead to singularities at the level of the metric. To avoid possible inconsistencies, we assume that $m>|a|$.

\begin{figure}
         \includegraphics[width=0.85\hsize]{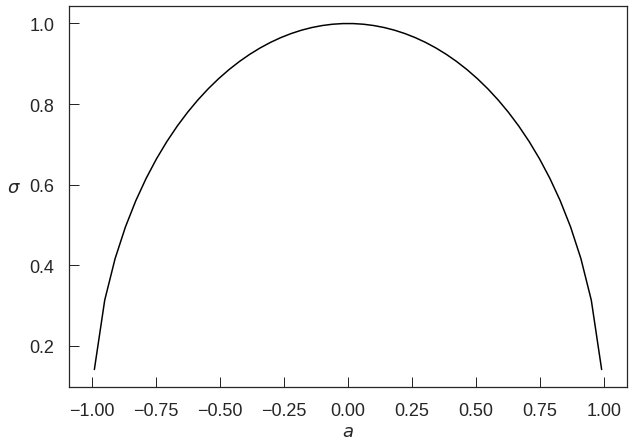}
         \includegraphics[width=0.85\hsize]{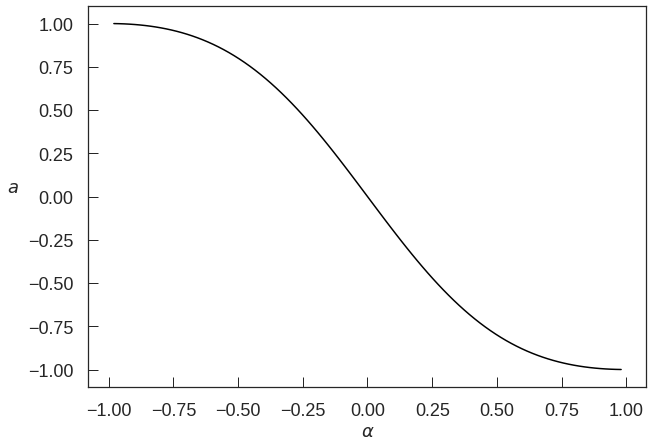}
         \caption{The rotation parameter $a$ as a function of the parameters $\sigma$ and $\alpha$ with $m=1$.}
    \label{sigmaa}
\end{figure}

\color{black}
It is convenient to consider $m$ and $a$ as the independent parameters, instead of $\sigma$ and $\alpha$. 
Indeed, by setting $a=0$, we obtain the limiting case of the static $\rm q-$metric. Moreover, for $\rm q=0$, the above solution is stationary with $a$ representing the specific angular momentum of the source. 
Figure \ref{sigmaa} shows the relationship of the rotation parameter $a$  with $\sigma$ and $\alpha$, respectively.

In general, the physical meaning of the parameters entering the metric can be clarified by calculating the relativistic multipole moments \cite{geroch1,geroch2,hansen,frutos2018relativistic}. We obtain


\be
M_0 = m + \mathcal{X}  \ , 
\ee
\be
J_1 = m a + 2 a \mathcal{X} \ ,
\ee
\be
M_2 = -m^3 + m\sigma^2 +(\frac{7}{3}\sigma^2-3m^2)\mathcal{X} -m \mathcal{X}^2 - \frac{1}{3} \mathcal{X}^3 \ ,
\ee
where $\mathcal{X}:=q\sigma$. All the higher moments can be expressed in terms of the above multipoles. 
Moreover, all the odd mass moments $M_{2k+1}$ and even angular momentum moments $J_{2k}$ vanish identically as a result of the existing reflection symmetry with respect to the equatorial plane.

Since all the higher multipole moments can be written in terms of the first and second multipoles, from the above expressions for the relativistic multipole moments, we see that the three parameters $m$, $\rm q$, and $a$ that enter the metric are present in the expressions for all multipoles through the quantity $\mathcal{X}$.



\begin{figure}
    \centering
    \includegraphics[width=\hsize]{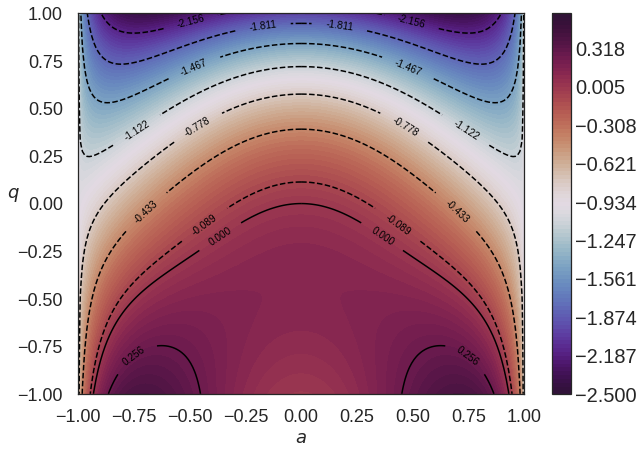}
    \caption{Values of the relativistic quadrupole $M_2$ for $m=1$ and different values of the rotation parameter $a$ and the quadrupole parameter $\rm q$.} 
    \label{fig:relativisticM}
\end{figure}

To illustrate the dependence of the multipoles from the independent parameters, we plot in Figure \ref{fig:relativisticM} 
the behavior of the relativistic quadrupole $M_2$. We see that $M_2$ is symmetric with respect to the value of $a$, but no symmetry exists with respect to the parameter $\rm q$. This is due to the fact that in this case the gravitational field does not depend on the direction of rotation, whereas the sign of quadrupole $\rm q$ determines the shape of the source, which can be either prolate or oblate, corresponding to different gravitational fields. 

\begin{figure}
   \centering
    \includegraphics[width=\hsize]{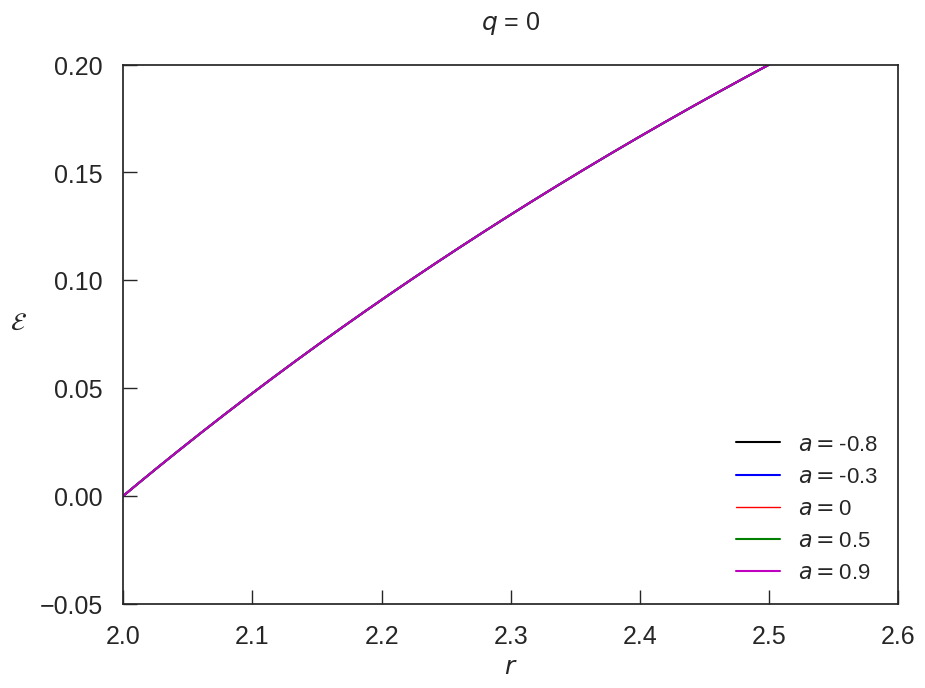}
   \caption{The Ernst potential on the equatorial plane for vanishing $\rm q$ and different spin parameters $a$.}
   \label{fig:ernstq0}
   \end{figure}

\begin{figure}
\centering
\includegraphics[width=0.9\hsize]{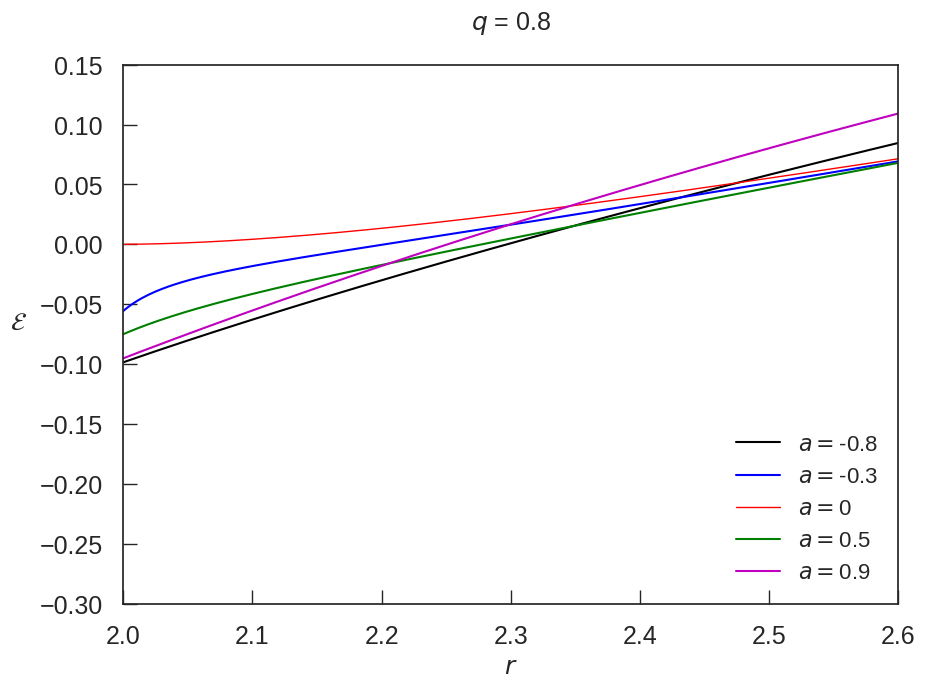}
 \includegraphics[width=0.9\hsize]{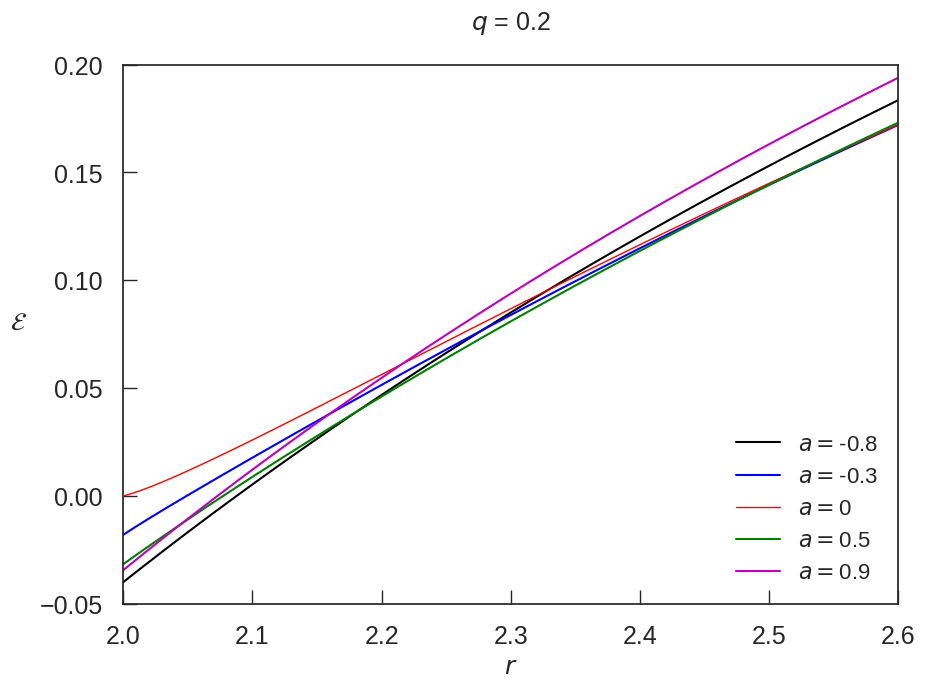}
\includegraphics[width=0.9\hsize]{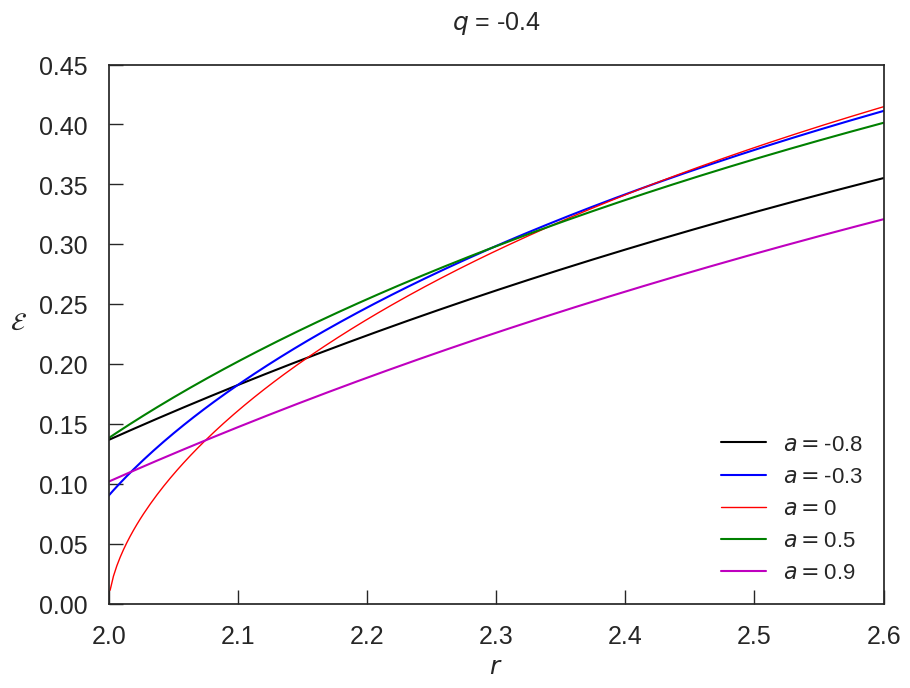} 
 \caption{The Ernst potential, $\mathcal{E}$, is depicted for various values of $\rm q$ and $a$}
 \label{ernst}
\end{figure}

 To further analyze the physical meaning of the stationary q-metric, we consider the corresponding Ernst potential that can be written as \cite{toktarbay2014stationary, frutos2018relativistic}
\begin{align}
    \mathcal{E}=\left(\frac{x-1}{x+1}\right)^ q \left[\frac{x-1+(x^2-1)^{-q}d_+}{x+1+(x^2-1)^{-q}d_-}\right],
\end{align}
where
\begin{align}
    d_{\pm}&=-\alpha^2(x\pm1)h_+h_-(x^2-1)^{-q}\nonumber\\
    +&i\alpha[y(h_+ + h_-)\pm (h_+ - h_-)],\
\end{align}    
and
\begin{align}
h_{\pm}&= (x\pm y)^{2q}.   
\end{align}
In Figs. \ref{fig:ernstq0} and \ref{ernst}, we study the behavior of the Ernst potential on the equatorial plane $\theta=\pi/2\ (y=0)$ as a function of the radial coordinate $r$ for different values of the two independent parameters $ \rm q$ and $a$.  

In the case 
$\rm q=0$, the Ernst potential reduces to 
\begin{align}\label{ernstq0}
    \mathcal{E}=\frac{a^2(x-1)- (\sigma-m)^2(x+1)}{a^2(x+1)- (\sigma-m)^2(x-1)}. 
\end{align}
since $|a| \in [0,1)$, its power two is a small quantity. Also, the coefficients of $a^2$ in the numerator and denominator of the Ernst potential \eqref{ernstq0} are of the same order and comparable, i.e., the values of $\mathcal{E}$  for different values of $a$ are very close to each other. The result is shown in Figure \ref{fig:ernstq0}. We can see that for each value of $a$, the Ernst potential is a continuous function of $r$ that tends monotonically to a constant at infinity. However, when we plot the Ernst potential for different values of $ q\neq 0$, as shown in Figure \ref{ernst}, we notice that there are certain points at which the  Ernst potentials for different values of $a$ coincide. Since the Ernst potential contains all the information about the gravitational field, we conclude that at the intersection points the corresponding spacetimes are identical. A detailed numerical 
analysis of the intersection points shows that they are located close to the radius value $r=2m$.

To clarify this point, we investigate the position of the curvature singularities of the stationary q-metric. In fact, for the static q-metric (\ref{qmetric}) it was shown in \cite{quev11} that there exist an exterior singularity located at $r=2m$ and an interior singularity with $r<2m$ and a shape that depends on the coordinate $\theta$ and on the value of the quadrupole parameter $\rm q$. 

\begin{figure*}
\begin{tabular}{cc}
     \centering
         \includegraphics[width=0.5\hsize]{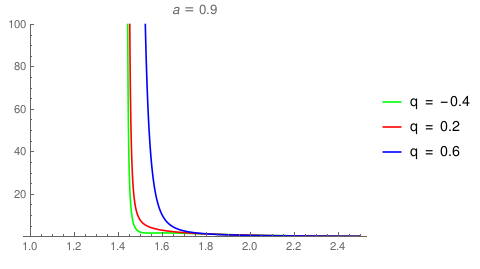}&
         \includegraphics[width=0.5\hsize]{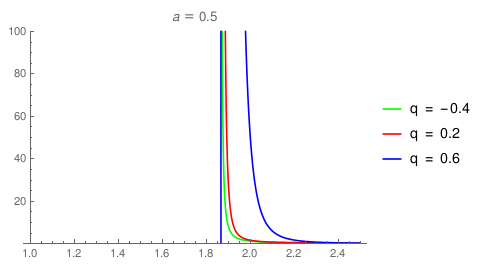} \\
          \includegraphics[width=0.5\hsize]{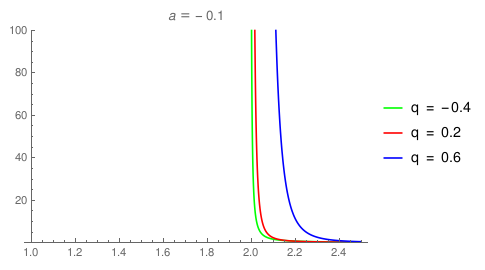}&
         \includegraphics[width=0.5\hsize]{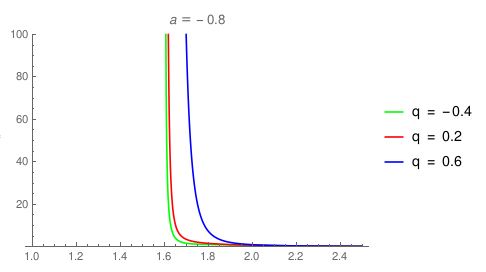}
  \end{tabular}               
   \caption{Behavior of the Kretschmann scalar on the equatorial plane as a function of the radial distance $r$ for different values of $\rm q$ and $a$.}
    \label{Kscalar}
\end{figure*}

One would expect that in the spacetime of the stationary q-metric, the singularity structure is affected by the presence of the rotation parameter $a$. To see this, we calculate 
 the Kretschmann scalar $\mathcal{K}=R_{\nu \mu \tau \theta}R^{\nu \mu \tau \theta}$ on the equatorial plane  and plot it in Figure \ref{Kscalar} for the values of $a$ and $\rm q$  that we will use later to construct the thick disk model. The plots show for the chosen values of the parameters that the singularities are always located around $r=2M$, i.e., the rotation changes only slightly the location of the static singularity. 
The same behavior is observed for other 
non-vanishing values of the quadrupole parameter within the range $q\in (-1,1]$.
 

  Moreover, we see that the location of the singularities is a decreasing function of $a$; namely, as $|a|$ increases, the singularities are concentrated closer to the value $r=2M$. 
 We conclude that the rotation only slightly affects the position of the curvature singularity, which is always located close to the hypersurface $r=2m$. 
 
 On the other hand, as mentioned above, there are  points in spacetime, where the stationary q-metric with different values of $\rm q$  and $a$ represent the same gravitational field. As shown above, these intersection points are always located very close to the hypersurface $r=2m$, i.e., where the spacetime becomes singular. 
  Since we are interested in studying thick disks that are far away from the hypersurface $r=2m$, the intersection points are not an obstacle for the investigation of the disks physical properties.

\begin{figure}
\centering
         \includegraphics[width=0.8\hsize]{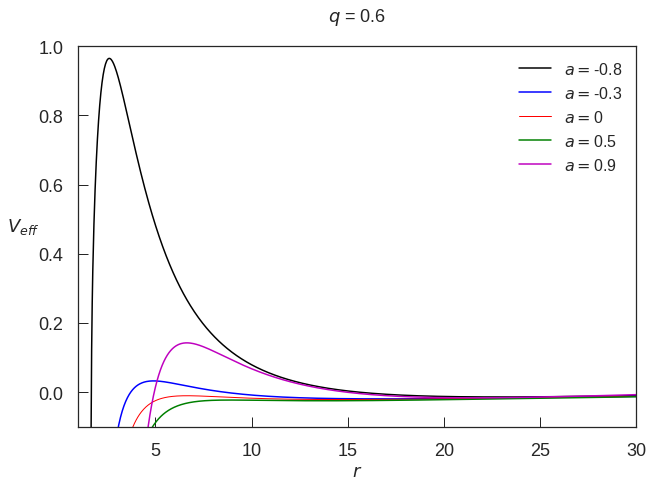}
         \includegraphics[width=0.8\hsize]{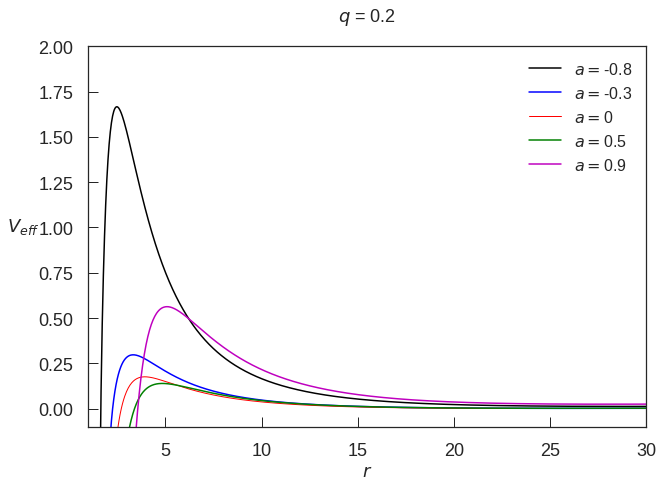}
         \includegraphics[width=0.8\hsize]{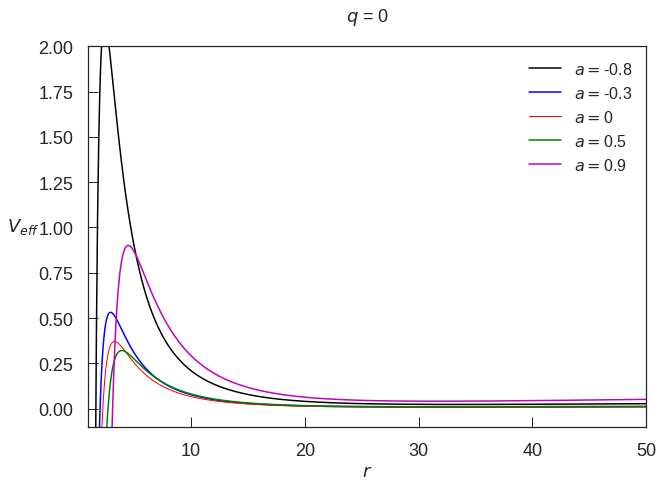}
         \includegraphics[width=0.8\hsize]{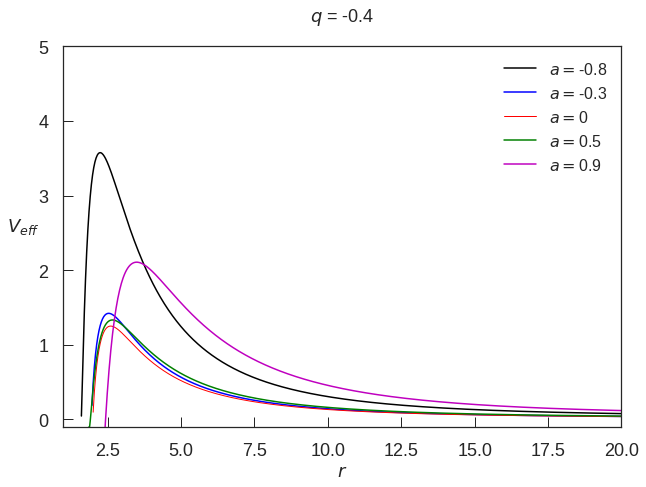}
       \caption{Effective potential $V_{\rm Eff}$ for different values of $a$ and $\rm q$.}\label{veff}
 \end{figure}

Furthermore, the analysis of the motion of test particles on the equatorial of the spacetime described by the stationary q-metric can be reduced to the analysis of the effective potential which in terms of the metric components can be expressed as 

\be
V_{\rm eff}= -1+\frac{E^2g_{\phi \phi}+E\ell g_{t\phi}-\ell^2g_{tt}}{g_{t\phi}^2+g_{tt}g_{\phi\phi}},
\label{effpot}
\ee
where $\ell$ and $E$ are the specific angular momentum and specific energy, respectively. In Figure \ref{veff}, we illustrate the behavior of the effective potential for different values of $\rm q$ and $a$ that we will use later in the construction of thick disks. 

\begin{figure*}
\centering
\begin{tabular}{cccc}
         \includegraphics[width=0.25\hsize]{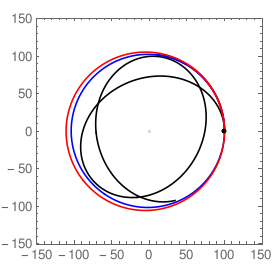}&
          \includegraphics[width=0.25\hsize]{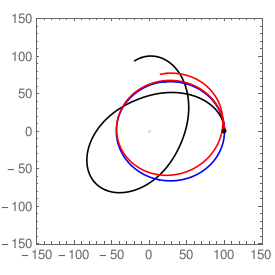}&
           \includegraphics[width=0.25\hsize]{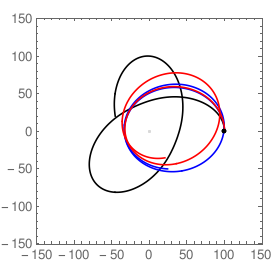}&
         \includegraphics[width=0.25\hsize]{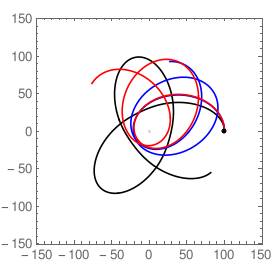}\\
         \includegraphics[width=0.25\hsize]{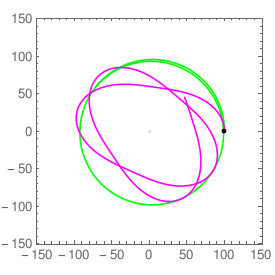}&
          \includegraphics[width=0.25\hsize]{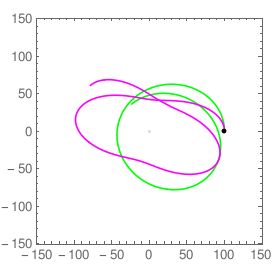}&
          \includegraphics[width=0.25\hsize]{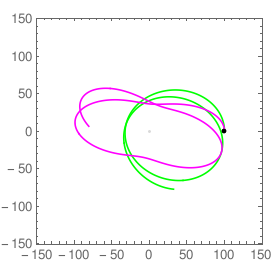}&
         \includegraphics[width=0.25\hsize]{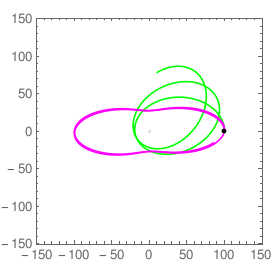}\\
   \end{tabular}                  
   \caption{Bounded trajectories for various combinations of $a$ and $\rm q$. The first row is dedicated to $a \leq 0$ and the second one to $a > 0$. The colors refers to the values of $a$ following the same code as the Figure \ref{bound}. From the left to the right $\rm q=-0.4$, $\rm q=0$, $\rm q=0.2$ and $\rm q=0.6$. All along the plots the following parameters remain constant: The angular momentum $L=8$, the initial point $r(0)=100$, and $ \phi(0)=0$.}
    \label{bound}
\end{figure*}

The analysis of the effective potential allows us to examine the regions where bounded orbits are possible to study the disk configuration. The result of this procedure is illustrated in Figure \ref{bound} for several values of the quadrupole $\rm q$ and the rotation $a$.


Our results show that indeed in the gravitational field of the stationary q-metric bounded orbits are allowed, which we can interpret as indicating the existence of accretion disks around the gravitational source.
Thus, in the next section, we explore the possibility of constructing models for thick disks on the background of the stationary q-metric.

\color{black}

\section{Profile of the toroidal disk}
\label{sec:tori}

The model of equilibrium tori for accretion discs is characterized by a negligible loss of mass and no self-gravity. 
In these models, the gravity plays
a crucial role in building the equipotential configurations. One of the important features of the tori model is that it exhibits locally a stabilizing mechanism against  thermal and viscous instabilities, and globally versus the Papaloizou and Pringle instability \cite{1981Natur.294..235A,1987MNRAS.227..975B}. In these configurations, the equation of state corresponds to that of a barotropic perfect fluid since the model is based  on the Boyer's condition. To briefly explain this model, we proceed as follows. The general stationary axisymmetric metric in the spherical-like coordinates is given by

\begin{align}
    {\rm d}s^2=g_{tt}{\rm d}t^2+2g_{t\phi}{\rm d}t {\rm d}\phi +g_{rr}{\rm d}r^2+g_{\theta \theta}{\rm d}{\theta}^2+g_{\phi \phi}{\rm d}{\phi}^2,
\end{align}
where $g_{\mu \nu} =g_{\mu \nu} (r,\theta)$. The stress-energy tensor  contains only the contribution of the perfect fluid; therefore, we have

\begin{align} 
T^{\mu}{}_{\nu}=wu_{\nu}u^{\mu}-\delta^{\mu}{}_{\nu}p,
   \end{align}
where $p$ is the pressure, $w$ is the enthalpy as measured by an
observer moving with the fluid, and the four velocity $u^{\mu}$ is given by $ u^{\mu}=(u^t,0,0,u^{\phi})$ since we assume that the fluid rotates in the azimuthal direction.

The relativistic Euler equation for the circular motion and $\Omega = \Omega(\ell)$ can be written as \cite{1978A&A....63..221A}


\begin{align}\label{maintori}
    \int_{p_{\rm in}}^{p}{\frac{{\rm d}p}{w}}=-\ln{\frac{|u_t|}{|(u_t)_{\rm in}|}}+\int_{\ell_{\rm in}}^{\ell}{\frac{\Omega{\rm d}\ell}{1-\Omega\ell}},
\end{align}
where the index $"\rm in"$ refers to the internal edge of the disk. In this case, the general relativistic version of the von Zeipel theorem is fulfilled. Accordingly, the surfaces of equal $\Omega$, $\ell$, $p$ and $w$ all are the same \cite{1978A&A....63..221A}. Hence, for the constant angular momentum distribution $\ell_0$, the total potential can be found as

\begin{align}\label{K33}
    W(r,\theta)=\frac{1}{2}\ln|\frac{g^{2}_{t \phi}-g_{tt} g_{\phi \phi}}{g_{\phi\phi}+2\ell_0g_{t\phi}+\ell^2_0g_{tt}}|,
\end{align}

\begin{figure}
         \centering
     \includegraphics[width=\hsize]{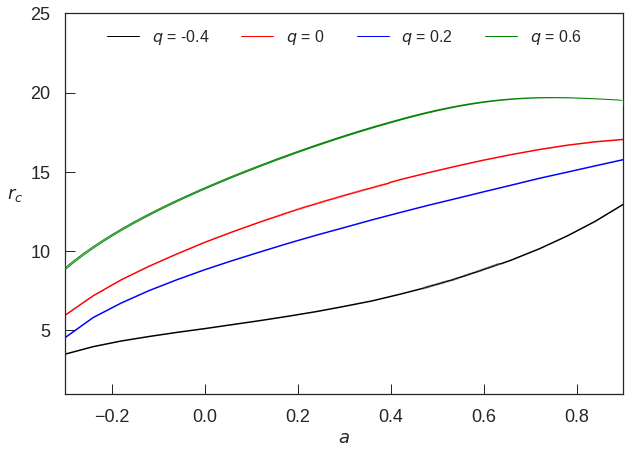}
    \caption{The variation of the position of the center of the disk as a function of $a$.} 
    \label{rc}
    \end{figure}

This model can be adjusted for both the constant and non-constant angular momentum distributions. For constant angular momentum profile that we adapted here, $W$ fulfills the following conditions \cite{1978A&A....63..221A},

\begin{align} \label{closed}
   \text{surfaces are}   \left\{
 \begin{array}{@{}ll@{}}
  \text{closed,} \quad \text{ if} \quad |\ell_{\rm ms}|<|\ell_0|< |\ell_{\rm mb}|, \\
  \text{open,} \quad \quad \text{if} \quad  |\ell_0|\geq |\ell_{\rm mb}|.
    \end{array}\right.
\end{align}
and if $|l_0|=|l_{\rm ms}|$, we end up with just a ring. Here mb and ms stays for marginally stable and marginally bound, respectively. Thus, for a perfect fluid matter rotating around an object described by the stationary $\rm q$-metric, the shapes and location of the equipressure surfaces follow from the specified angular momentum distribution $\ell$. In this work, we assume $p \ll \rho$ meaning  $w \sim \rho$, and the equation of state is given by
\begin{equation}
    p=K\rho^{\frac{4}{3}}
\end{equation}
where $K$ is a constant. In addition, the cusp point is located at the smallest radius of $\ell_0$ and the Keplerian angular momentum intersections, whereas  the biggest one characterizes the center of the disk. This radius is shown in Figure \ref{rc} as a function of $a$ for different values of $\rm q$. We see that $r_c$ is an increasing function of $a$ and $\rm q$. It means that for any fixed value of $\rm q$ (or $a$), for larger values of $a$ (or $\rm q$) the disk can be constructed farther away from the central object. In addition, the plots of $r_c$ as a function of $a$ have a maximum for positive $\rm q$s and a minimum for negative $\rm q$s. This means that for positive values, if one continuously increases $a$, the disk configuration shifted from the central source and then come closer. It could be either an interesting behavior of this spacetime, or some signal to discard very large positive values of $\rm q$. This analysis is beyond the scope of the current work because it would imply a deeper investigation of the physical properties of the background metric.

\begin{figure}
         \centering
  \includegraphics[width=\hsize]{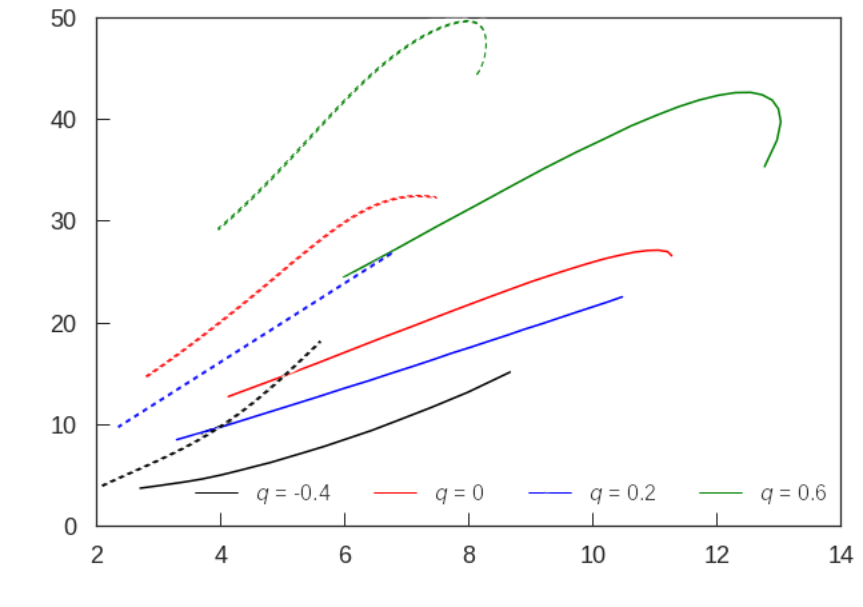}
    \caption{The variation of $\ell_{mb}^2$ (dashed line)  as a function of $r_{\rm mb}$, and the variation of $\ell_{ms}^2$ (thick line) as a function of $r_{\rm ms}$. Besides, the places of $r_{\rm mb}$ and $r_{\rm ms}$ also change with varying $\rm q$ and $a$. From the left to the right for each curve, $a$ goes from negative to positive values.}
    \label{fig:lmblms}
    \end{figure}

In Figure \ref{fig:lmblms}, we plot the square of the specific angular momentum at the marginally stable $r_{\rm ms}$ and marginally bound $r_{\rm mb}$ radii for different values of $\rm q$ and $a$.  As mentioned earlier, to have closed equipressure surfaces \eqref{closed}, we need to choose $\ell_0$ between the two curves for each model.

In fact, the angular momentum profiles $\ell^2_{\rm mb}$ and $\ell^2_{\rm ms}$ and their orbits $r_{\rm mb}$ and $r_{\rm ms}$, as shown in Figure \ref{fig:lmblms}, have the same behavior as $r_c$ with respect to parameters $a$ and $\rm q$. 

In addition, the area between these two profiles also increases, leading to the appearance of a wider region, where closed equipotential surfaces can exist because the chosen angular momentum $\ell_0$ should be within this region. 
A consistent approach for choosing this profile for different values of parameters consists in fixing the constant angular momentum a $\ell_0^2=\frac{1}{2}(\ell_{mb}^2+\ell_{ms}^2)$ for all the models.

\begin{figure}
\centering
         \includegraphics[width=0.9\hsize]{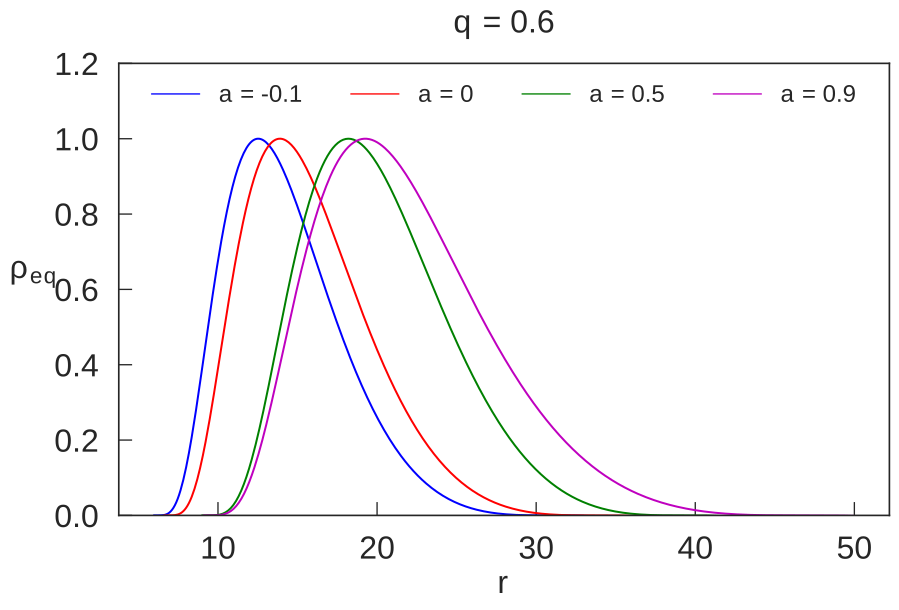}
         \includegraphics[width=0.9\hsize]{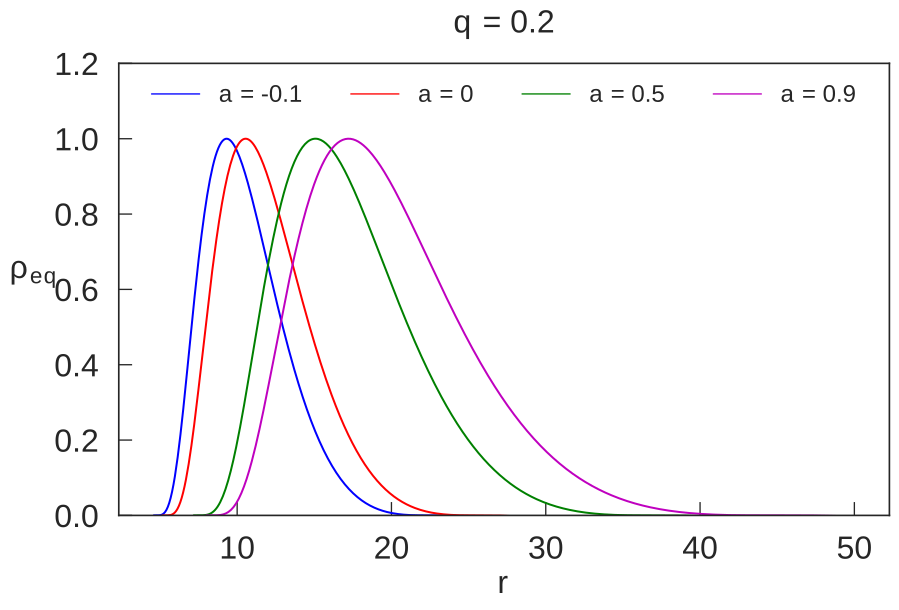}
          \includegraphics[width=0.9\hsize]{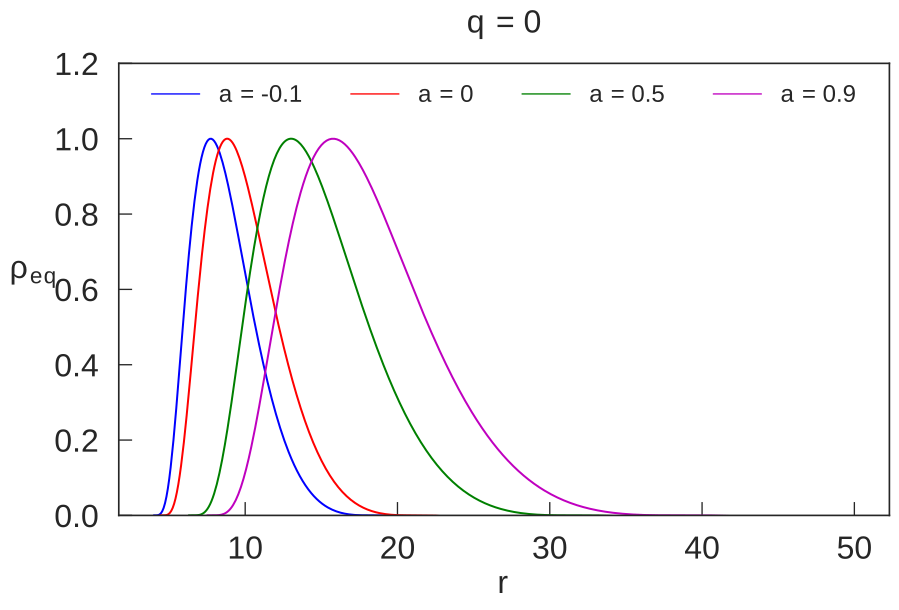}
         \includegraphics[width=0.9\hsize]{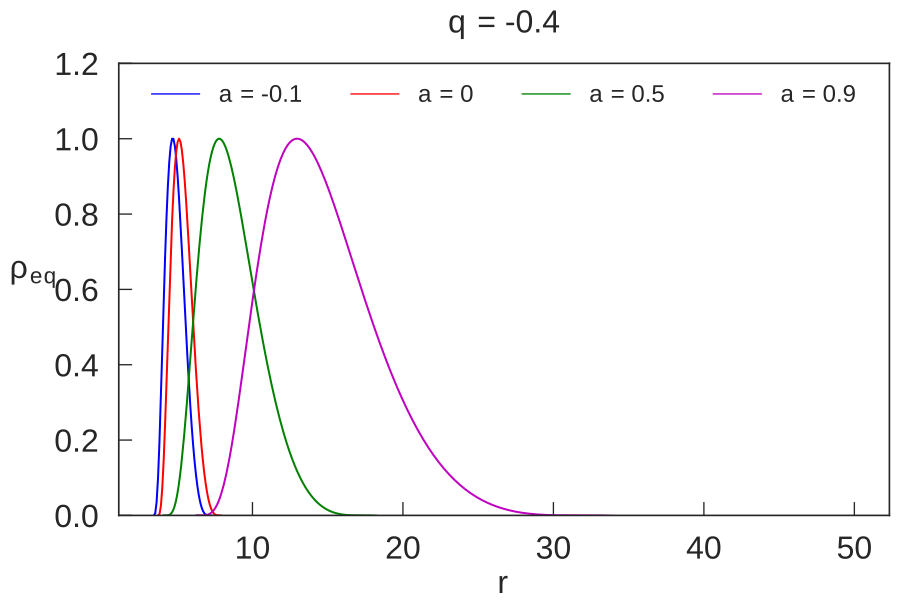}

   \caption{Density profile on the equatorial plane for the solutions given in  Figure \ref{fig:rho}. The plots are scaled at each case by the value at the center.}
    \label{RhoEq}
\end{figure}

To see the impact of the parameter $a$, in Figure \ref{RhoEq}, we plot the rest-mass density profiles on the equatorial plane for different values of $a$ and each chosen value for $\rm q$. We see that as $\rm q$ or $a$ increases, the maximum of the rest-mass density shifts farther away from the central object, which is consistent with the results of Figures \ref{rc} and \ref{fig:lmblms}.

\begin{figure*}
\centering
\begin{tabular}{cccc}
       \includegraphics[width=0.25\hsize]{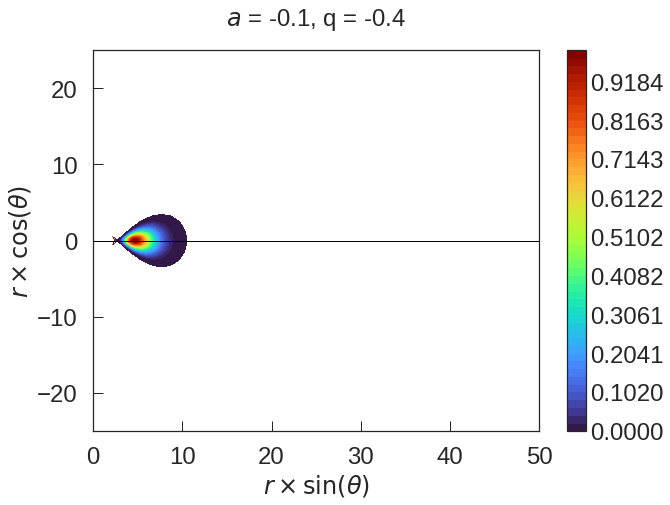}&
      \includegraphics[width=0.25\hsize]{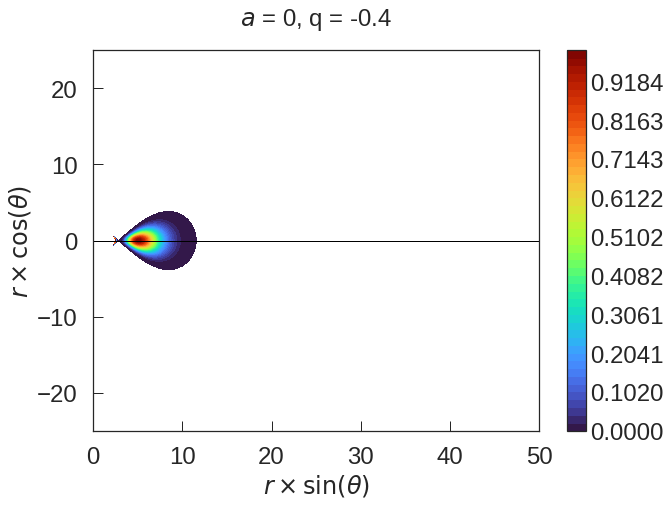}&
      \includegraphics[width=0.25\hsize]{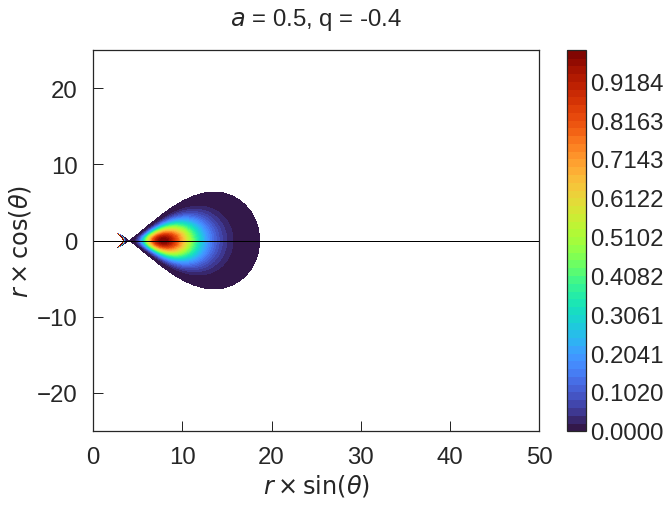}&
     \includegraphics[width=0.25\hsize]{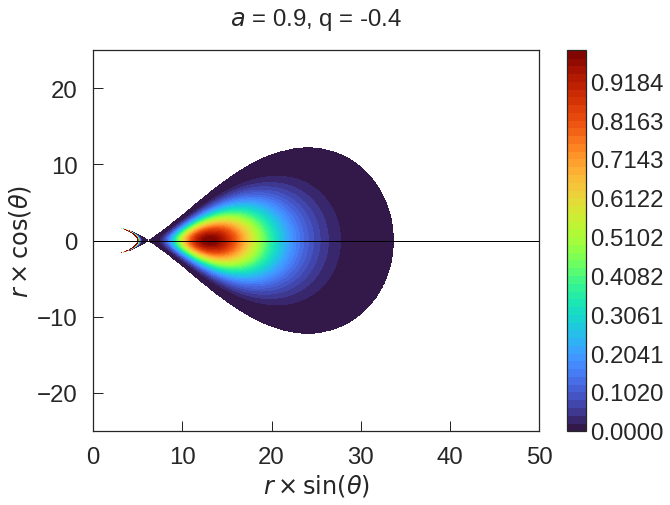}\\
         \includegraphics[width=0.25\hsize]{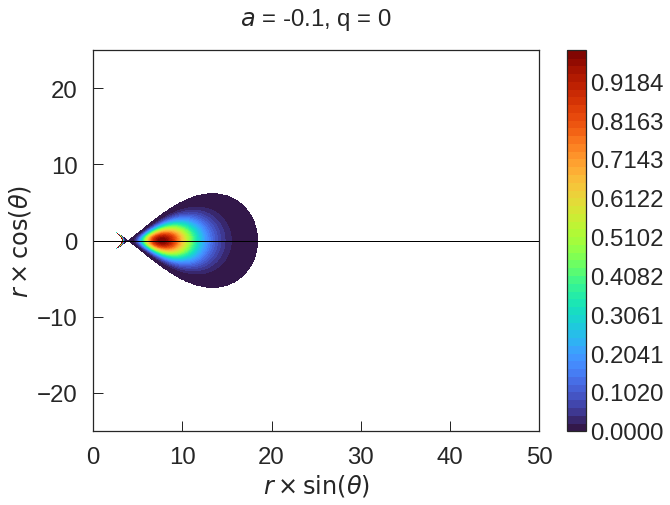}&
          \includegraphics[width=0.25\hsize]{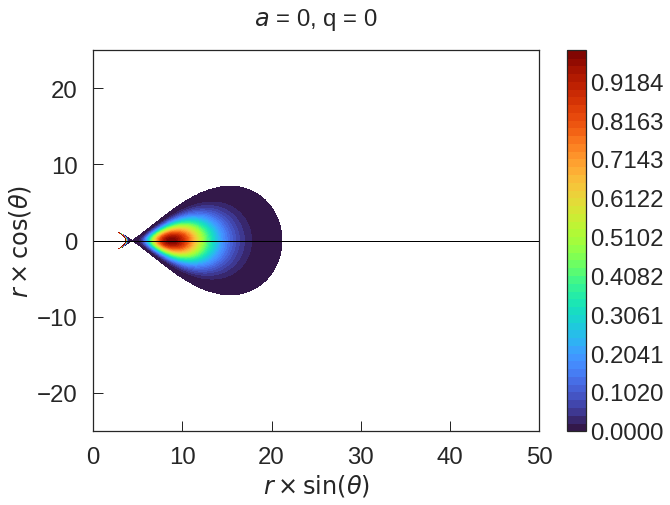}&
        \includegraphics[width=0.25\hsize]{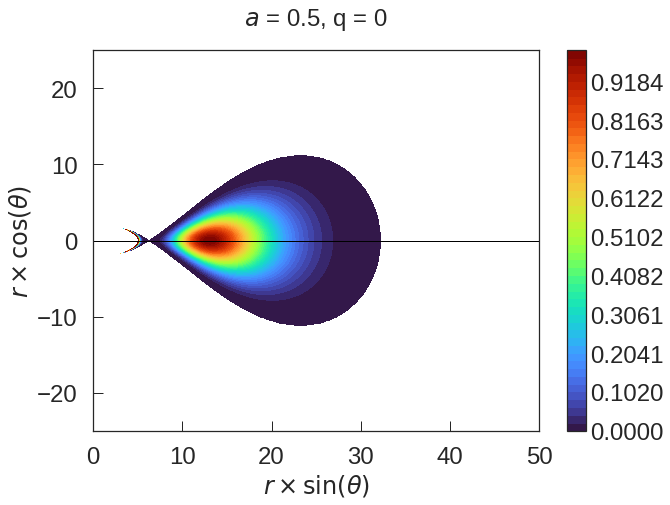}&
         \includegraphics[width=0.25\hsize]{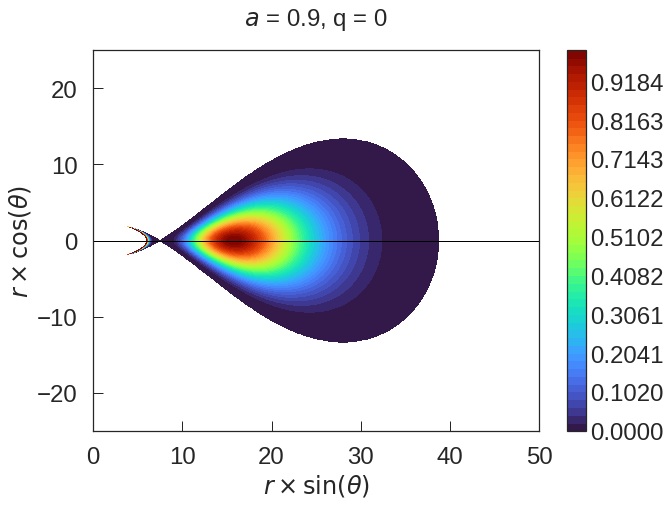}\\
         \includegraphics[width=0.25\hsize]{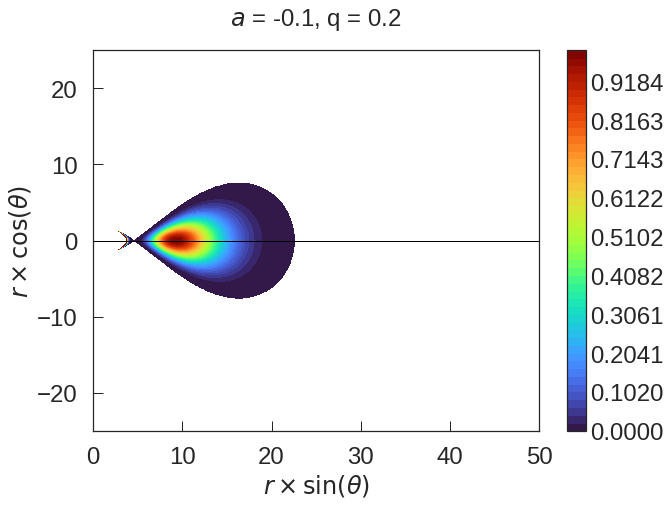}&
          \includegraphics[width=0.25\hsize]{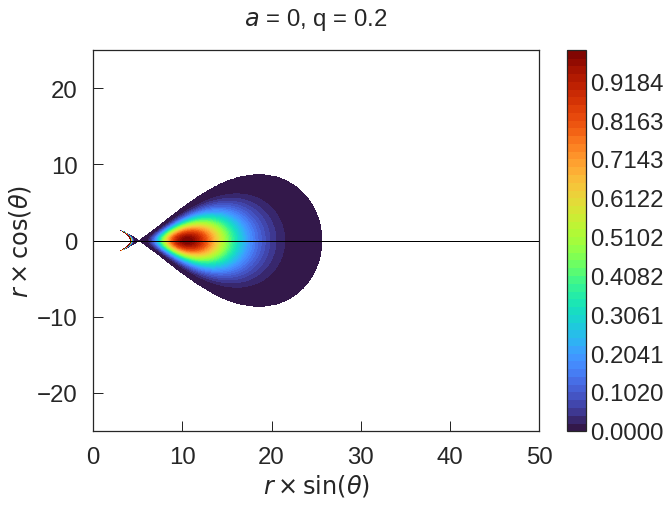}&
         \includegraphics[width=0.25\hsize]{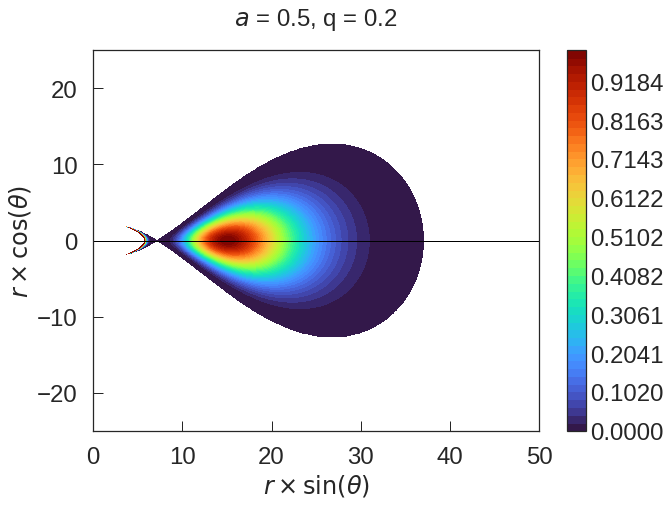}&
         \includegraphics[width=0.25\hsize]{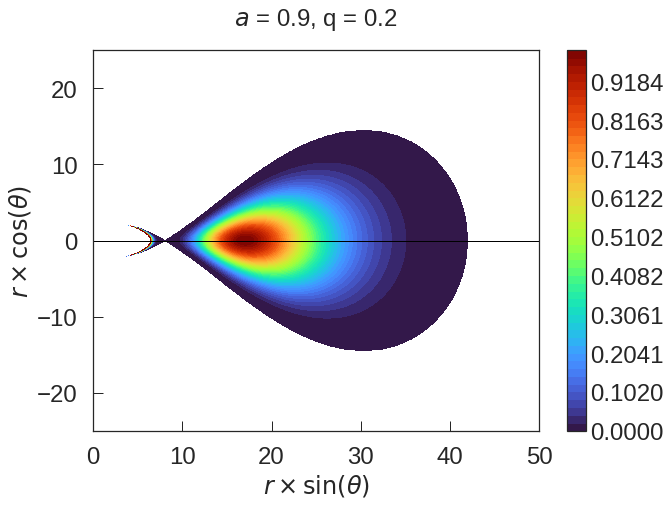}\\
         \includegraphics[width=0.25\hsize]{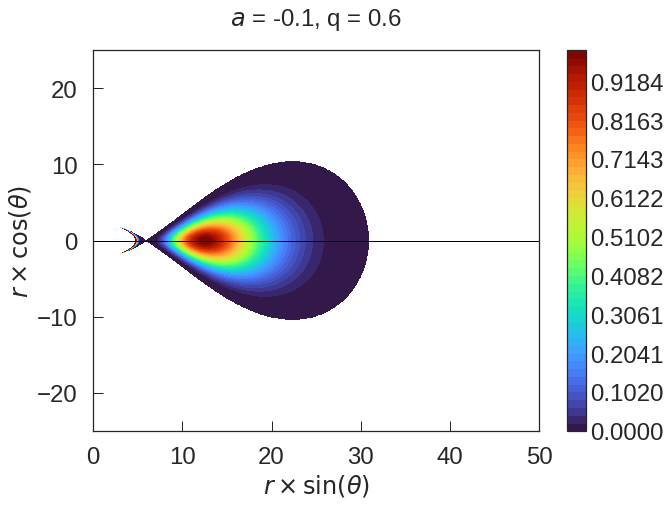}&
          \includegraphics[width=0.25\hsize]{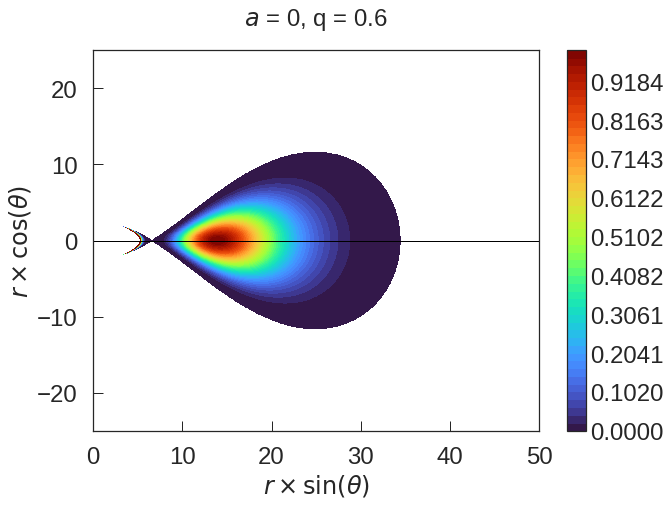}&
         \includegraphics[width=0.25\hsize]{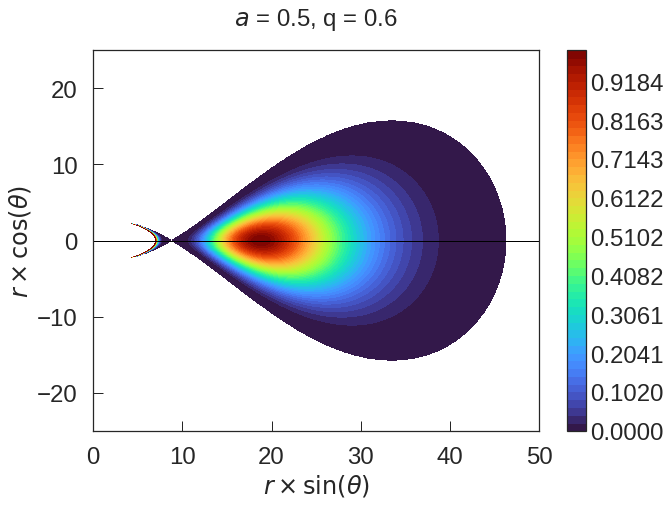}&
         \includegraphics[width=0.25\hsize]{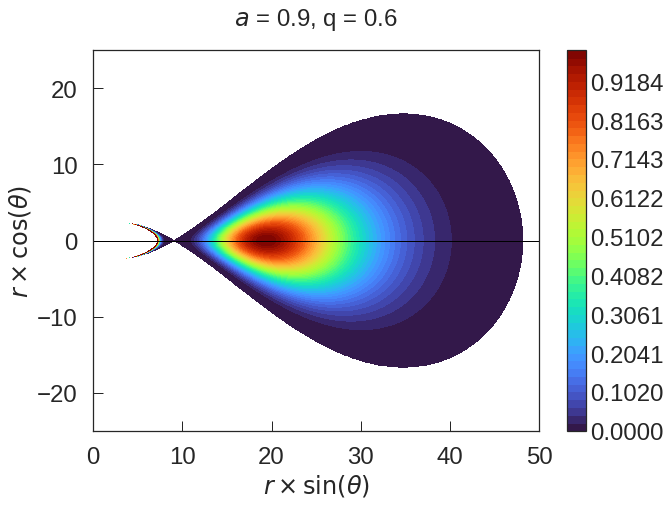}\
   \end{tabular}                  
   \caption{Equidensity surfaces for various values of $\rm a$ (vertical direction) and $\rm q$ (horizontal direction). All along the plots, $\ell_0^2=(\ell_{mb}^2+\ell_{ms}^2)/2$, with the $\ell_{mb}^2$ and $\ell_{ms}^2$ values associated to $\rm a$ and $\rm q$.}
    \label{fig:rho}
\end{figure*}

To study the influence of the parameters $\rm q$ and $a$ on the morphology of the equipressure surfaces, in Figure \ref{fig:rho}, the disk structure is plotted for the same values that were used previously. In the second column, we chose $a=0$ for the static $\rm q$-metric and compare it with the stationary case.

In Figure \ref{fig:rho}, a comparison among the different rows tells us about the role of the quadrupole $\rm q$ for any chosen $a$, while a contrast among the columns shows the influence of the parameter $a$ on the model for any chosen value of $\rm q$. Clearly, one can see that by increasing the value of $a$, the disk moves away from the compact object. Besides,  the disk shape becomes extended in the radial direction. 
A similar pattern is obtained for larger values of $\rm q$s. Then, both parameters seem to have the same effect on the disk structure. However, a deeper analysis reveals that the growth rate caused by increasing $a$ is higher than by increasing $\rm q$. For the case $a=0$ the result is in  good agreement with the results found in \cite{2021A&A...654A.100F} for a constant angular momentum. Furthermore, in the second column and row, we have the Schwarzschild case and we see that by changing these parameters, the configuration of the disk smoothly deviate from this particular choice $a=0=\rm q$. 

In conclusion, with respect to the central source, the size and position of the disk  are a monotonically increasing functions of both parameters $a$ and $\rm q$. Therefore, it seems reasonable by considering the properties and the overall structure of a thick accretion disk, one may estimate the rotation and quadrupole parameters of the central compact object at least to find upper and lower bounds on them.

\section{Conclusions and summary}\label{sec:con}

In this paper, we analyzed the motion of particles in the stationary $\rm q$-metric considering relatively small quadrupole $\rm q$.

We also investigated the structure of  thick disks on this background to reveal the  physical properties of this spacetime. 
In particular, we studied the shape and properties of the equipressure surfaces of the thick disk model. In addition, we compare the results of our analysis with the limiting cases of the static q-metric and the  Schwarzschild metric. 

In the first part of this work, we analyzed the properties of  the Ernst potential as well as the effective potential of the stationary q-metric on the equatorial plane. Interestingly, it turns out that there are certain points on the equatorial plane at which the Ernst potential has the same value for different values of $a$, indicating a sort of degeneracy of the gravitational field at the intersection points. However, an analysis of the Kretschmann scalar shows that the intersection points are located in a region very close to the curvature singularities, which are always around the hypersurface $r=2M$. This degeneracy does not affect the structure of the accretion disks that are always located far away from the singularity.

In the second part, we explore the structure of thick tori in the spacetime of the stationary q-metric. In general, we conclude that the parameters $\rm q$ and $a$ drastically affect the shape of the disk. Indeed, the larger the values of $\rm q$ and $a$, the larger the disk configuration,  with an extended shape along the radial direction, and the farther away from the central object is the disk located. These results show that, in principle, it should be possible to determine the rotational and quadrupole parameters of the central object by measuring the shape and location of accretion disks.  

It would be interesting to continue the investigation of this spacetime by studying the behavior of other astronomical systems that could exist in the gravitational field of the stationary q-metric. In particular, to test the applicability of this solution in the numerical setups and compare it with other backgrounds.

\newpage

\section{Acknowledgements}
S.F. thanks the Cluster of Excellence EXC-2123 Quantum Frontiers - 390837967 and the research training group GRK 1620 "Models of Gravity",  founded by the German Research Foundation (DFG). A.T. thanks the research training group GRK 1620 "Models of Gravity", funded by DFG.
The work of H.Q. was partially supported by  UNAM-DGAPA-PAPIIT, Grant No. 114520, and  Conacyt-Mexico,  Grant No. A1-S-31269.

\bibliographystyle{unsrt}
\bibliography{sqmain}

\end{document}